\documentclass[11pt]{article} 
\usepackage{rotating} 
\usepackage{amsbsy} 
\usepackage{pifont} 
\usepackage{amssymb} 
 
\renewcommand\baselinestretch{1.10} 
\def\ket#1{ $ \left\vert  #1   \right\rangle $ } 
\def\ketm#1{  \left\vert  #1   \right\rangle   } 
\def\spr#1#2{ $ \left\langle #1 \left\vert \right. #2 \right\rangle $ } 
\def\sprm#1#2{  \left\langle #1 \left\vert \right. #2 \right\rangle   }

\def\tmm#1{      \buildrel - \over #1  }   
   
\def\tpm#1{      \buildrel + \over #1  }   

\begin{document} 
 
\title{Spectroscopic $LSJ$ notation for atomic levels obtained \\[0.1cm]
       from relativistic calculations} 
 
\author{G.\ Gaigalas$\, ^{a,b}$\footnote{Email: gaigalas@mserv.itpa.lt},  
        T.\ Zalandauskas$\, ^b$ and  
        S. Fritzsche$\, ^a$                                          \\ 
        \\ 
        ${}^a$ Fachbereich Physik, Universit\"a{}t Kassel,           \\ 
               Heinrich--Plett--Str. 40, D--34132 Kassel, Germany.   \\ 
        \\ 
        ${}^b$ Vilnius University Research 
	Institute of Theoretical Physics and Astronomy,              \\ 
        A.\ Go\v{s}tauto 12, Vilnius 2600, Lithuania.                } 
 
\maketitle 
 
\date{} 
 
\begin{abstract} 
Today, relativistic calculations are known to provide a very successful 
means in the study of open--shell atoms and ions. But although accurate 
atomic data are obtained from these computations, they are traditionally 
carried out in $jj-$coupling and, hence, do often not allow for a simple
$LSJ$ classification of the atomic levels as needed by experiment.
In fact, this lack of providing a proper spectroscopic notation from
relativistic structure calculations has recently hampered not only the 
spectroscopy  of medium and heavy elements, but also the interpretation and 
analysis of inner--shell processes, for which the occurrence of 
additional vacancies usually leads to a very detailed fine structure.

Therefore, in order to facilitate the classification of atomic levels from 
such computations, here we present a program 
(within the \textsc{Ratip} environment) which help transform the atomic wave
functions from $jj-$coupled multiconfiguration Dirac--Fock computations 
into a $LS-$coupled representation.
Beside of a proper $LSJ$ assignment to the atomic levels, the program also 
supports the full transformation of the wave functions if required for
(nonrelativistic) computations.
\end{abstract}

\newpage 
 
{\large\bf PROGRAM SUMMARY} 
 
\bigskip

{\it Title of program:} \textsc{Lsj}. 
 
\bigskip

{\it Catalogue number:} ADTL
 
\bigskip 
 
{\it Program summary URL:} http://cpc.cs.qub.ac.uk/summaries/ADTL
 
\bigskip  
 
{\it Program obtainable from:} CPC Program Library,  
     Queen's University of Belfast, N. Ireland.  
 
\bigskip

{\it Licensing provisions:} None. 
 
\bigskip

{\it Computer for which the new version has been tested:}   IBM RS 6000, 
     PC Pentium III.                                        \newline 
     {\it Installations:} University of Kassel (Germany).   \newline 
     {\it Operating systems:} IBM AIX 4.1.2+, Linux 7.1.+ 
      
\bigskip

{\it Program language used in the new version:} ANSI standard Fortran 90/95. 
 
\bigskip

{\it Memory required to execute with typical data:} Memory requirements 
     depend on the shell structure and the size of the wave function expansion  
     which is used to represent the atomic levels. 
 
\bigskip

{\it No.\ of bits in a word:}  All real 
     variables are parametrized by a \texttt{selected kind parameter} and,  
     thus, can easily be adapted to any required precision as supported  
     by the compiler. Presently, the \texttt{kind} parameter is set to 
     double precision (two 32--bit words) in the module 
     \texttt{rabs\_{}constant}. 
 
\bigskip

{\it CPU time required to execute test data:}  5 seconds on a 1 GHz  
                                               Pentium III processor. 
 
\bigskip

{\it Distribution format:} compressed tar file.  
 
\bigskip


 
\bigskip

{\it Nature of the physical problem:}  \newline 
     The spectroscopic $LSJ$ notation is determined for atomic levels 
     which were calculated previously in the framework of the $jj-$coupled  
     multiconfiguration Dirac--Fock (MCDF) model. This notation is based on  
     a complete $jj-LS$ transformation of the leading $jj-$coupled 
     configuration state functions (CSF) in the wave function representation 
     of the selected levels. 
      
\bigskip

{\it Restrictions onto the complexity of the problem:}  \newline  
     The $jj \rightarrow LS$ transformation of the ($jj-$coupled) CSF
     is supported for all shell structures including open $s-,\, p-,\, d-,$ 
     and $f-$shells. For shells with $l > 3$ (i.e.\ beyond the $f-$subshells), 
     however, a proper transformation of the antisymmetrized 
     subshell states can be carried out only for the case of one or two  
     \textit{equivalent} electrons. This restriction also applies for the 
     transformation of the $g_{\,7/2}$ and $g_{\,9/2}$ subshell states which 
     are otherwise supported by the \textsc{Ratip} package 
     [S.~Fritzsche, J.~Elec.\ Spec.\ Rel.\ Phen.\  
     \textbf{114--116} (2001) 1155.]. 
     The $jj \leftrightarrow LS$ transformation matrices, which are applied
     internally by the program, are consistent with the definition of the
     (reduced) coefficients of fractional parentage [G.~Gaigalas, Z.~Rudzikas and C.~Froese Fischer, 
         At.~Data  Nucl.~Data Tables, \textbf{70} (1998)~1;
	 G.~Gaigalas, S.~Fritzsche and Z.~Rudzikas, 
         At.~Data  Nucl.~Data Tables, \textbf{76} (2000) 235.] as published
     previously.

\bigskip

{\it Unusual features of the program:}  \newline 
     The \textsc{Lsj} program is designed as a part of the \textsc{Ratip} 
     package      [S.~Fritzsche, J.~Elec.\ Spec.\ Rel.\ Phen.\  
     \textbf{114--116} (2001) 1155.] for the computation of (relativistic) atomic transition and 
     ionization properties. This (new) component therefore supports the 
     transformation  of all atomic states which are generated either with 
     \textsc{Ratip} or by means of the \textsc{Grasp92} code 
     [F.~A.\ Parpia, C.~F.\ Fischer, and I.~P.\ Grant, 
     Comput. Phys. Commun.\ \textbf{94} (1996) 249.]. 
     Moreover, the  normalization of the transformed states is tested within
     the $LSJ-$coupled basis.

%
%
%

\newpage 
 \vspace*{2.5cm} 
 
{\large\bf LONG WRITE--UP} 
 
\bigskip

\section{Introduction} 

In atomic spectroscopy, the standard $LSJ$ notation of the atomic levels 
is frequently applied for classifying the low--lying level structures of 
(most) atoms and ions. The knowledge of 
the basic $LSJ$ symmetries of the levels 
often helps the interpretation of atomic spectra without detailed 
computations. In fact, the $LSJ$ spectroscopic 
notation is found useful also in order to explain the excitation and the decay
properties of individual levels, at least qualitatively. Since, in addition,
most atomic structure calculations were carried out within a $LS-$coupled 
basis in the past, the $LSJ$ designation of the levels is taken today as the 
\textit{standard notation} in the literature as well as in the most data bases.

\medskip

In recent years, however, relativistic structure computations also 
became feasible and were frequently performed, when accurate data were 
required. Based on either the Dirac--Coulomb or Dirac--Coulomb--Breit 
Hamiltonian, for instance, these computations are traditionally carried out in 
$jj-$coupling, from which the $LSJ$ notation cannot be derived so easily, especially
if atomic configurations with (several) open shells occur. 
In the last few years, therefore, the lack of a proper spectroscopic 
notation deriving from relativistic calculations has hampered not only further
analysis of open $d-$ and $f-$shell elements but also the study and 
interpretation of the inner--shell processes where the occurrence of (additional)
inner--shell holes often result in a very detailed fine--structure. 
In order to obtain the standard notation in such a case, an additional 
transformation of the atomic states (from the $jj-$coupled into a $LS-$coupled 
basis) is needed first before the full benefit is drawn from such a 
relativistic treatment. 
But although several tools and tabulations are available today for carrying 
out the transformation between different coupling schemes, there is no
program available which would directly support the well--known relativistic 
structure codes such as \textsc{Grasp92} \cite{Parpia/CFF/Grant:96} or 
\textsc{Ratip} \cite{Fritzsche:01,Fritzsche/CFF/Dong:00,FritzscheFG:02}. 
The reasons for this lack are simple: (i) the rapidly increasing 
complexity of the transformation matrices, when several open shells are involved, 
and (ii) no simple access to the standard quantities from the 
atomic shell model.

\medskip

Apart from the explicit $jj-LS$ transformation of the many--electron basis, 
there is a second method known from the literature 
\cite{SandarsB:65,Armstrong:66,Armstrong:68} which helps to derive the $LSJ$ 
classification of atomic levels from relativistic computations. This method 
applies the \textit{(effective)} relativistic operators within an $LS-$coupled 
basis and, hence, can be directly implemented into the non--relativistic 
theory, i.e.\ by using the familiar Pauli spinors for the construction of the
many--electron basis. So far, however, 
these effective operators have neither been derived in detail for all the 
physical cases of interest \cite{KarazijaJ:01} nor does a full implementation 
exist which is competitive with the standard relativistic 
codes. Moreover, the use of effective operators within a Pauli spinor basis 
cannot account for the different \textit{radial} behaviour of equivalent 
electrons with $ j \,=\, l \,\pm\, 1/2  \;\, (l \,\ge\,1)$ as is found to be important, 
in particular, for the behaviour of inner--shell electrons in medium and 
heavy elements. 
Therefore, the explicit $jj-LS$ transformation of the many--electron basis 
seems to us a much simpler route to derive the $LSJ$ designation of the
atomic states, making use of the well--known relativistic structure 
codes such as \textsc{Grasp92} \cite{Parpia/CFF/Grant:96} or \textsc{Ratip} 
\cite{Fritzsche:01} as its input. This $jj-LS$ transformation of the basis
states has recently been generalized also to allow an arbitrary number 
of open shells in the construction of the symmetry--adapted functions
\cite{Gaigalas/RZ:HCI02}.

\medskip

To facilitate the classification of the atomic levels and wave functions 
from the relativistic calculations, here we present the \textsc{Lsj} program as 
a new component of the \textsc{Ratip} package \cite{Fritzsche:01}. 
This component supports both the $LSJ$ classification of the level structure 
of open--shell atoms and ions as well as the (complete) transformation of 
the corresponding wave functions. 
But the program may also be of value in solving the problems of a more general
character, where the transformation from the $jj$-- into $LS$--coupling scheme 
and
vice versa is performed, e.g. a transformation of this type is needed when
one seeks the optimal coupling scheme after calculating the energy spectra
in a non--relativistic or relativistic approximation. But the problems of this
type are quite independent of those investigated here. They are related only
to a similar transformation being made. Therefore the authors have not aimed at
creating a general program in the frame of the \textsc{Ratip} package that would
be of use while the physical problems are formulated in the most general case,
especially with certain risks originating from such a formulation of a problem
when the transformation considered is based on the transformation 
coefficients~\cite{Gaigalas/ZR:02}.
The values of these depend on the phase convention that is chosen to obtain the
spin--angular coefficients and on the definition of the coefficients of
fractional parentage (CFP). Of course, the program presented in this work may serve as
a basis for solving the other problems in atomic physics. But the authors have
not aimed at the exhaustive description of the program on purpose, because in
that case the technical way of answering those questions is not as important as
is the methodological one, related to the phase convention and the definition of
CFP. The program is based on the methodology that uses the Fano--Racah
convention, and the CFPs are defined via the quasispin 
formalism~\cite{Gaigalas/RF:97}. The CFPs are
published in~\cite{Gaigalas/RF:98,Gaigalas/SR:00}.
In all other cases this transformation leads to erroneous
results. Therefore, before starting to perform any tasks related to the use of
the program in other packages or trying to solve any problems of other type, we
recommend that the authors of the paper should be contacted.

\medskip

In the following section, we first provide 
a brief outline on the transformation of the atomic states in going from 
the $jj-$coupled to a $LS-$coupled many--electron basis. 
This shows how the $LSJ$ spectroscopic notation 
is obtained as a symmetry of the leading term in the wave function
representation within the $LS-$coupled basis; however, we leave out most of the
details concerning the construction and recoupling of the symmetry functions 
in the atomic shell model since that material has been presented recently 
\cite{Gaigalas/Fri:02}. In section 3, we describe the program 
organization and how the code is distributed. 
To demonstrate the interactive use of the \textsc{Lsj} 
program, section 4 displays --- by two examples --- the derivation of the 
$LSJ$ notation for the low--lying excited levels of beryllium as well as 
for the ground--state of gadolinium. Finally, the conclusions and a brief 
outlook are given in section 5.

\section{Theoretical background} 

The task to derive the $LSJ$ designation of atomic levels from relativistic
calculations can be summarized rather easily without much knowledge 
about the atomic shell model. Having, for example, a representation 
of an atomic state $\psi$ with respect to a $jj-$coupled symmetry basis, 
the (first few) leading terms in the wave function expansions need to be 
transformed into a $LS(J)-$coupled basis from which, then, the proper 
spectroscopic notation is given by the symmetry of (again) the leading term. 
At the same time, the contributions from the subsequent terms in 
the $LS-$coupled basis indicate the admixture of other $LS$ symmetries and, 
hence, the deviation from a pure $LS$ or Russell--Saunders coupling scheme. 
Of course, the explicit $jj-LS$ transformation of the atomic state functions 
(ASF) requires quite detailed information about the construction and the 
coupling of the symmetry adapted basis functions, both in $jj-$ as well as 
$LS-$coupling. For most practical applications, however, these details are 
irrelevant. 

\medskip

In this write--up of the \textsc{Lsj} program, we therefore restrict the 
discussion mainly to the representation of the atomic states in the various 
coupling schemes, but otherwise refer the reader to the literature
\cite{Gaigalas/Fri:02,Rudzika-book}. While this restriction enables us to 
explain the basis steps of a  $jj-LS$ transformation of atomic states 
from standard 
relativistic computations, and how these steps are implemented within the 
program, it avoids the need of providing too many details about the definition
of symmetry--adapted functions and about the theory of transformation in the
atomic shell model.

\subsection{Representation of atomic states} 

In principle, of course, any valid coupling scheme can be used in order
to represent the (many--electron) wave functions in atomic structure
computations. In relativistic calculations, however, a $jj-$coupled basis 
appears particularly appropriate as the symmetry of the Dirac equation for a
single electron leads naturally to solutions which have a well--defined 
total angular momentum, $ \mathbf{j} \,=\, \mathbf{l} \,+\, \mathbf{s}$, 
but no defined projections (onto the $z-$axis) of either the 
orbital angular momentum $\mathbf{l}$ nor the spin $\mathbf{s}$ of the electron. 
In the multiconfiguration Dirac--Fock (MCDF) or configuration interaction (CI)
methods, these one--electron solutions of the Dirac equation (which 
are finally determined for some effective potential) are then used also for 
the set--up of the many--electron basis functions, the so--called 
configuration state functions (CSF). Owing to the overall symmetry of atoms 
with respect to a rotation or inversion of the coordinates, these CSF are 
constructed in order to have a well--defined total angular momentum $J$ 
and total parity $P$. For a given set of such configuration states, 
$\{ \ketm{\gamma_r JP}, \;\, r = 1...n_c \}$, the atomic states are obtained 
simply as a linear combination
\begin{eqnarray}  
\label{Psi-jj}  
   \ketm{\Psi_\tau (J^P)} & = &  
   \sum_{r}^{n_c} \, a_r^{\,(jj)} (\tau) \, \ketm{\gamma_r JP}  
\end{eqnarray} 
of CSF of the same symmetry, where $\tau \,=\, 1,\ 2,\ ... $ enumerates the 
atomic levels (of the given symmetry) and $\gamma_r$ denote the set of all 
further quantum numbers as required for an unique classification of the 
symmetry--adapted basis. Obviously, these quantum numbers and the details 
of their coupling play an important role in the transformation of the basis
from one coupling scheme into another. For the present discussion, however, 
it is sufficient to say that the CSF basis in (\ref{Psi-jj}) spans a (finite) 
part of the $N-$electron Hilbert space and that the $a_r^{\,(jj)} (\tau)$  
represent the (usual) Fourier coefficients for the expansion of an atomic state
with respect to this basis.  As seen from ansatz (\ref{Psi-jj}), moreover,
only $J$ and $P$ are in fact good quantum numbers which, apart from the 
subshell angular momenta (see below) and the total energy of the ASF, can 
be used for the classification of the atomic levels.

\medskip

An (almost) equivalent\footnote{Since a $jj-$coupled basis allows, due to 
distinctions of \textit{equivalent} electrons with $j \,=\, l \,\pm\, 1/2
\; (l \,\ge\,1)$, a somewhat greater flexibility in the representation of 
the atomic states, 
a \textit{full} equivalence of the representations in $jj-$ and $LS-$coupling 
is obtained only in the limit of a complete, infinitely 
large basis.}
but alternative representation of the ASF can be obtained within a 
$LS-$coupled basis
\begin{eqnarray}  
\label{Psi-LS}  
   \ketm{\Psi_\tau (J^P)} & = &  
   \sum_{s}^{n_d} \, a_s^{\,(LS)} (\tau) \, \ketm{\gamma_s\ LS\ JP}  \, ,
\end{eqnarray}
where one--electron spin--orbital functions with well--defined orbital 
angular momentum $l$ (as well as with projections $m_l$ and $m_s$ of the 
orbital angular momentum and the spin onto the quantization axis) are used 
in the set--up of the corresponding CSF basis. Instead of a defined total 
angular momentum $j$ of each single electron, however, here the orbital angular 
momenta and the spins of all electrons are first coupled together separately
in order to obtain a total orbital angular momentum,
$ \mathbf{L} \,=\, \mathbf{l}_1 \,+\, \mathbf{l}_2 \,+\, ...$, and a total
spin, $ \mathbf{S} \,=\, \mathbf{s}_1 \,+\, \mathbf{s}_2 \,+\, ...$. The 
symmetry $J$ of the CSF with respect to a rotation of the coordinates
are obtained then from the coupling 
$ \mathbf{J} \,=\, \mathbf{L} \,+\, \mathbf{S}$ of the total orbital and spin
angular momenta. In $LS-$coupling, therefore, $L$ and $S$ are (approximately)
also \textit{good} quantum numbers which can be used for the classification 
of the atomic states. And, in fact, such an $LS$ coupling of the atomic levels
appears to be well realized in Nature, in particular for light atoms and 
ions with, say, $Z \lessapprox 20$ as well as for the low--lying levels of 
most neutral atoms.

\medskip

As pointed out, the two representations (\ref{Psi-jj}) and (\ref{Psi-LS}) 
of the atomic states are almost equally suitable to determine the energies and
properties of an atom. To derive the $LSJ$ designation of
an atomic state, for which a representation $\{ a_r^{\,(jj)} \}$ in 
$jj-$coupling is given, it is sufficient to expand each CSF 
\ket{\gamma_r JP} into a $LS-$coupled basis and to sum up the contributions
of the different CSF, using the proper weight $ a_r^{\,(jj)} (\tau)$. For two
orthonormalized sets of CSF, such a transformation 
\begin{eqnarray}  
\label{LS-jj-csf}  
   \ketm{\gamma_r\ LS\ JP} \;\longleftrightarrow\;   
   \ketm{\gamma_s JP} \,  
\end{eqnarray}  
is represented by an \textit{orthogonal} matrix. The knowledge of this matrix is
therefore all that is needed in practice in order to carry out the 
transformation explicitly and to find the symmetry of the leading term in 
the $LS-$coupled basis. The \textsc{Lsj} program enables the user to evaluate
and to apply this matrix for any (given) configuration basis in $jj-$coupling
which is used for the description of open--shell atoms and ions. There are,
however, a few restrictions on the set--up of these transformation matrices
which are reasonable to \textit{maintain} for practical computations.
The next subsection explains these restrictions and summarizes the steps 
which are needed to construct the transformation matrix.

\subsection{Transformation of configuration symmetry functions} 

For identical fermions, the symmetry--adapted CSF must represent 
many--electron functions which are also antisymmetric under the interchange 
of any pair of electron coordinates. In the atomic (and nuclear) shell model, 
therefore, it is common practice to construct the CSF from antisymmetrized 
(sub--)shell states which describe $N$ \textit{equivalent} electrons in a 
given shell. In $LS-$coupling, for example, such subshell states for the 
shell $(nl)$ are written as~\cite{Rudzika-book} 
\begin{equation} 
\label{eq:quasispin-LS-a} 
   \ketm{nl^{\,N} \; \alpha \nu LS} \, ,
\end{equation} 
where $n$ is the principal quantum number and $\alpha $ represents all 
those quantum numbers which, in addition to the total orbital angular 
momentum $L$, total spin $S$ and seniority number $\nu$, are needed for 
the unique classification of 
these states. In practice, an additional number $\alpha $ is required only 
for shells with orbital angular momenta $l \ge 3$, i.e.\ for electrons 
from the $f$--, $g$--, \ldots shells. A more detailed discussion  of all the
possible subshell states (for shells with $l \le 3$) has been given recently
in Refs.\ \cite{Gaigalas/SR:00,Gaigalas/Fritzsche/Fricke:01}.
The CSF with $u$ shells in $LS-$coupling takes the form
\begin{eqnarray} 
\label{eq:quasispin-LS-general} 
\lefteqn{
   \ketm{\gamma\ LS\ JP} }
   \nonumber \\[1ex]
   & \equiv &
   \ketm{n_1l_1^{\,N_1} \; n_2l_2^{\,N_2} ... n_ul_u^{\,N_u} \; 
   \alpha_1 \nu_1 L_1S_1 
   ... \alpha_u \nu_u L_uS_u
   \; L_{12}S_{12} ... L_{u-1 \, u}S_{u-1 \, u}
   \; LS \; J} \, .
\end{eqnarray} 
The \textit{additional} intermediate angular momenta 
$L_{i \, j}, \; S_{i \, j}$
will arise in (\ref{eq:quasispin-LS-general}) if the 
subshell states of two or more open shells 
are coupled to each other as needed for CSF with more than a single open shell.

\medskip 
 
In $jj-$coupling, similarly, the subshell states of $N$ equivalent electrons 
of a subshell $(n\kappa)$ are represented by \cite{Gaigalas/Fri:02}
\begin{equation} 
\label{eq:quasispin-jj-a} 
   \equiv \ketm{n\kappa^{N} \; \alpha \nu J} \;  
\end{equation} 
where $\kappa$ is the relativistic (angular momentum) quantum number  
\begin{eqnarray} 
   \kappa & = & \pm \, (j+1/2) \qquad \mbox{for} \qquad l \;=\; j \pm 1/2 \;  
\end{eqnarray} 
and where two further quantum numbers $\nu$ and $J$ are found sufficient 
in order to classify all subshell states with $j$ = $1/2$, $3/2$, $5/2$, $7/2$, 
and $9/2$ unambiguously. 
An additional number $\alpha $ is required only 
for subshell states with angular momenta $j \ge 9/2$. 
Again, Refs.\ 
\cite{Gaigalas/SR:00,Gaigalas/Fritzsche/Fricke:01} provide a more thorough
discussion and a complete list of all allowed states for (relativistic)
subshells with $j \,\le\, 9/2$. Of course, the transformation of the 
subshell states lies at the heart of any transformation which is to be 
carried out for the symmetry--adapted CSF and which need to be solved first.
The $jj \leftrightarrow LS $ transformation of the subshell states 
(in both directions) was therefore the topic of a recent extension of the 
\textsc{Racah} package \cite{Gaigalas/Fri:02,Fritzsche:97} which was 
utilized also for the implementation of the present program. 
In going from a $jj-$ to a $LS-$coupled basis, however, there are two
difficulties in the transformation procedures which one must keep in mind. 
These difficulties are related (i) to the group--theoretical differences 
of the subshell states in $jj-$ and $LS-$coupling, respectively, as well as 
(ii) to the re--coupling of the antisymmetrized 
subshell states which is generally required in order to bring these
states into a pre--defined coupling order on both sides of the 
transformation matrix elements. We will briefly discuss, in turn, these two 
issues to make the reader aware also of the limitations of the present program.

\medskip

Although, in $LS-$ and $jj-$coupling, the basic notations 
(\ref{eq:quasispin-LS-a}) and (\ref{eq:quasispin-jj-a}) of the subshell states 
appear to be rather similar, these states generally belong to different 
irreducible representations of the $SO_3$ rotation group because, 
in $jj-$coupling, 
each (nonrelativistic) $nl$--shell can be `split' into two 
(relativistic) subshells with total angular momenta $j_{\pm} \,=\, l \pm 1/2 $.
In the expansion of a single shell state \ket{nl^{\,N} \alpha L S} into a 
$jj-$coupled basis, therefore, both subshell states with $j_{-}$ and $j_{+}$,
i.e.\  \ket{\tmm{\kappa}^{\tmm{N}} \tmm{\nu} \tmm{J}} and
       \ket{\tpm{\kappa}^{\tpm{N}} \tpm{\nu} \tpm{J}}, will occur. Here, again,
we make use of the relativistic quantum number $\kappa$ to simplify 
the notation below. Obviously, $ N \,=\, \tmm{N} + \tpm{N} $ and  
$\tmm{\kappa} \,=\, -(\tpm{\kappa}+1) \,>\,0$ must apply where the notation 
$\,\tmm{\kappa} \;>\,0$ and $\,\tpm{\kappa} \;<\,0$ has been chosen to be
consistent with the corresponding total $j_{\pm} \,=\, l \pm 1/2 $ and 
subshell angular momenta.

\medskip

Making use of this notation, the transformation between the $jj-$ and
$LS-$coupled subshell states takes the form
\footnotesize
\begin{eqnarray} 
\label{Matrixjj-LSDef-one-b} 
   \ketm{(\tmm{\kappa}^{\tmm{N}} \tmm{\nu} \tmm{J},\ 
          \tpm{\kappa}^{\tpm{N}} \tpm{\nu} \tpm{J})\ J}  
   & = & 
   \sum_{\alpha LS} \, 
   \ketm{l^{\,(\tmm{N}+\tpm{N})} \; \alpha LS\ J} 
   \sprm{l^{\,(\tmm{N}+\tpm{N})} \; \alpha LS\ J}{ 
         (\tmm{\kappa}^{\tmm{N}} \tmm{\nu} \tmm{J},\ 
          \tpm{\kappa}^{\tpm{N}} \tpm{\nu} \tpm{J})\ J} 
\end{eqnarray} 
\normalsize 
including a summation over $\alpha$ as well as the total subshell orbital 
angular momentum $L$ and the subshell spin $S$. As is clear from our notation 
in Eq.\
(\ref{Matrixjj-LSDef-one-b}), 
\ket{(\tmm{\kappa}^{\tmm{N}} \tmm{\nu} \tmm{J}, \ 
      \tpm{\kappa}^{\tpm{N}} \tpm{\nu} \tpm{J})\ J} 
represents a coupled state with well--defined total angular momentum
$J$ which is built from the corresponding $jj-$coupled subshell states with  
$j_{\pm} \,=\, l \pm 1/2 $ and the total subshell angular momenta $\tmm{J}$ 
and $\tpm{J}$, respectively. This expansion of the $jj-$coupled subshell states
therefore shows that, for any number of open (sub--)shells in the construction
of the CSF, the transformation of these symmetry functions can always be 
traced back to the transformation coefficients 
\small 
\begin{eqnarray} 
\label{subshell-coefficients}
   \sprm{l^{\,(\tmm{N}+\tpm{N})} \; \alpha LS\ J}{ 
         (\tmm{\kappa}^{\tmm{N}} \tmm{\nu} \tmm{J},\ 
          \tpm{\kappa}^{\tpm{N}} \tpm{\nu} \tpm{J})\ J}  
\end{eqnarray} 
\normalsize 
of the individual subshell.
And, in fact, these (Fourier) coefficients are the `building blocks' for any
transformation of the atomic states into a $LS-$coupled basis. They are
implemented explicitly into the \textsc{Lsj} component (by means of a recent
revision of the \textsc{Racah} package \cite{Gaigalas/Fri:02}) since their 
computation requires a recursive --- and rather expansive --- procedure which 
traces back the subshell states on both sides of expression 
(\ref{subshell-coefficients})
to their parents states with $N-1$ electrons, and so on.
The transformation coefficients (\ref{subshell-coefficients}) are implemented
for all (sub--) shells with $l \le 3$ and occupation numbers 
$ N \,=\, 1,\ 2,\ ...,\ 2l+1$; for all higher (allowed) occupation numbers
$ N \,=\, 2l+2,\ ...,\ 4l+2$, these coefficients are simply obtained by using
the \textit{electron--hole} symmetry and the tabulated coefficients for 
$ N' \,=\, 4l+2-N$ \cite{Gaigalas/Fri:02}. 

\medskip

A second difficulty in the $jj-LS$ transformation of atomic states arises from
the (re--) coupling of the antisymmetrized subshell states. Since, in general,
CSF with several open shells occur very frequently, a pre--defined coupling 
order of the individual subshell states 
$n_1 l_1^{\,N_1},\ n_2 l_2^{\,N_2},\ ...$
need first to be specified, such as
\begin{eqnarray} 
\label{LS-CSF} 
   \ketm{(...(((l_1^{\,N_1} \alpha_1 L_1 S_1, \,  
                l_2^{\,N_2} \alpha_2 L_2 S_2) L_{12}S_{12}, \, 
                l_3^{\,N_3} \alpha_3 L_3 S_3) L_{123}S_{123})...)\ J} 
\end{eqnarray} 
for an $LS-$coupled basis. Of course, a similar coupling order of the subshell
states could be written also for a $jj-$coupled basis
$ \ketm{\kappa_1^{\,N_1} \; \nu_1 J_1},\ 
  \ketm{\kappa_2^{\,N_2} \; \nu_2 J_2},\ ... $ 
where any (predefined) sequence would give rise to a valid many--particle 
basis. But the predefined sequence of $jj-$coupled subshell states also decides
how many (re--) coupling steps are needed to bring both sides in the overall
transformation matrix \spr{\gamma_s\ LS\ JP}{\gamma_r JP} into the same 
coupling order. For this reason, we request the $jj-$coupled CSF in the
\textsc{Lsj} program to be given in a (so--called)  \textit{standard order} 
\footnotesize 
\begin{eqnarray} 
\label{standard-order} 
   \ketm{(...(((((\tmm{\kappa}_1^{\tmm{N}_1} \tmm{\nu}_1 \tmm{J}_1,         \, 
             \tpm{\kappa}_1^{\tpm{N}_1} \tpm{\nu}_1 \tpm{J}_1) J_1,         \, 
             \tmm{\kappa}_2^{\tmm{N}_2} \tmm{\nu}_2 \tmm{J}_2) J^{\prime}_{12},\, 
             \tpm{\kappa}_2^{\tpm{N}_2} \tpm{\nu}_2 \tpm{J}_2) J_{12} 
             \tmm{\kappa}_3^{\tmm{N}_3} \tmm{\nu}_3 \tmm{J}_3) J^{\prime}_{123},\, 
             \tpm{\kappa}_3^{\tpm{N}_3} \tpm{\nu}_3 \tpm{J}_3) J_{123}...)J} 
\end{eqnarray} 
\normalsize 
in which these functions have to fulfill two additional conditions: 
 
\smallskip 
     
(i) If both subshells with common $l_i$, i.e.\ $ \tmm{\kappa}_i$ and  
    $ \tpm{\kappa}_i$, appear in the expansion, these two subshells always 
    occur successively in the sequence 
    $ (\tmm{\kappa}_i^{\tmm{N}_i} \tmm{\nu}_i \tmm{J}_i, \, 
       \tpm{\kappa}_i^{\tpm{N}_i} \tpm{\nu}_i \tpm{J}_i) \ J_{i} $. 
    Formally, we can use this sequence even for subshell states 
    with zero occupation if we interpret  
    $\ketm{\kappa^0 \nu=0\ J=0} \,\equiv\, 1$; in this case, the full 
    Clebsch--Gordan expansion for the symbolic expression 
    (\ref{standard-order})
    remains valid due to the orthonormality properties of the Clebsch--Gordan 
    coefficients. 
     
\medskip 
     
(ii) For the $jj-LS$ transformation of configuration states 
\small 
\begin{eqnarray} 
\label{general-trans}
   & & \hspace*{-0.8cm} 
   \left\langle 
    (...(((l_1^{\,N_1} \alpha_1 L_1 S_1, \,  
           l_2^{\,N_2} \alpha_2 L_2 S_2) L_{12}S_{12}, \, 
           l_3^{\,N_3} \alpha_3 L_3 S_3) L_{123}S_{123})...)\ J     
   \left\vert \right. 
    (...(((((\tmm{\kappa}_1^{\tmm{N}_1} \tmm{\nu}_1 \tmm{J}_1,         \, 
             \tpm{\kappa}_1^{\tpm{N}_1} \tpm{\nu}_1 \tpm{J}_1) J_1, \right. 
   \nonumber \\[0.2cm]  
   & & \hspace*{3.0cm} 
   \left. 
             \tmm{\kappa}_2^{\tmm{N}_2} \tmm{\nu}_2 \tmm{J}_2)J^{\prime}_{12}, \, 
             \tpm{\kappa}_2^{\tpm{N}_2} \tpm{\nu}_2 \tpm{J}_2) J_{12} 
             \tmm{\kappa}_3^{\tmm{N}_3} \tmm{\nu}_3 \tmm{J}_3)J^{\prime}_{123},\, 
             \tpm{\kappa}_3^{\tpm{N}_3} \tpm{\nu}_3 \tpm{J}_3) J_{123}...)\ J 
   \right\rangle 
\end{eqnarray} 
\normalsize 
    we further assume in \textit{standard order} that 
    $ l_1 \,=\, \tmm{\kappa}_1 \,=\, -(\tpm{\kappa_1}+1),\   
      l_2 \,=\, \tmm{\kappa}_2 \,=\, -(\tpm{\kappa_2}+1),\  ...$, i.e.\ 
    that the sequence of (sub--)shell states is the 
    \textit{same on both sides} of the transformation matrix. 
     
\medskip

For such a standard order, explicit expression for the $jj-LS$ transformation
matrix of the CSF with two open shells in $LS-$ or up to four open shells 
in $jj-$coupling are displayed in Ref.\ \cite{Gaigalas/Fri:02}.
With the present \textsc{Lsj} program, we now extend these expressions
to allow for any number of open subshells.

\medskip

Although, of course, the standard order sets a restriction onto the set--up 
of the $jj-$coupled CSF basis, it is the order which is typically applied 
in all relativistic structure codes such as \textsc{Grasp92} 
\cite{Parpia/CFF/Grant:96} or \textsc{Ratip} \cite{Fritzsche:01}. 
But even if a \textit{non--standard} order has been used, the 
\textsc{Lsj} component can still be utilized with only minor additional effort.
Since, in fact, the coupling of the subshell states is independent of 
the radial wave functions, it is usually sufficient to (re--) generate 
the CSF basis in 
standard order and just to repeat the diagonalization of the Hamiltonian matrix.
Of course, the representation $\{ a_r^{\,(jj)} (\tau) \}$ of the atomic 
states in (\ref{Psi-jj}) must always be consistent with the basis itself. 
Moreover, if in the CSF basis the occupied shells appear in a different 
sequence than the usual (non--relativistic) one $1s,\, 2s,\, 2p,\, 3s,\, ...$,
the condition of the standard order is nevertheless fulfilled  and the 
\textsc{Lsj} program can be applied without further modifications. In this case,
however, the order of the \textit{parent} states may become different 
and may finally result in a different $LSJ$ classification, if the parent 
states are accounted for in the overall designation of the atomic levels.

\section{Program organization and distribution} 

Although \textsc{Lsj} is provided as an independent program for the
transformation of atomic levels and wave functions from a $jj-$coupled 
into a $LS-$coupled basis, it is designed also as a new program component 
of the \textsc{Ratip} package 
\cite{Fritzsche:01,Fritzsche/CFF/Dong:00,FritzscheFG:02} and, hence, follows
design principles similar to the other components of this package.
During recent years, \textsc{Ratip} has been found to be a powerful tool for
studying the \textit{(relativistic atomic) transition and ionization 
properties} of atoms and highly--charged ions within rather arbitrary
configurations. Apart from the transition probabilities and lifetimes,
for incorporating the interaction of atoms with the radiation field, 
\textsc{Ratip} now includes further components which help calculate Auger 
and photoionization properties as well as the coherence transfer in atomic
cascades. As described previously, the \textsc{Ratip} package is built on 
\textsc{Grasp92} \cite{Parpia/CFF/Grant:96} which is used to generate the 
(relativistic) wave functions within the MCDF model.

\medskip

However, here we shall not recall much about \textsc{Ratip} itself, neither 
about its overall structure nor about previous applications, since this has 
been described previously \cite{Fritzsche:02}. Apart from a few further 
applications to incorporate atomic processes with one electron in the 
continuum, our recent concerns pertained to a long life--cycle of the code, 
making use of an object--oriented design and, in particular, the concepts 
of Fortran 90/95. In fact, all components of \textsc{Ratip} are now well 
prepared to deal with large wave function expansions (\ref{Psi-jj}) as 
needed for the accurate handling of open--shell systems.

\subsection{Definition of derived data types} 
 
The proper choice of \textit{derived data types} certainly belongs to the key
features for a long--term maintenance and the development of a code. In
\textsc{Ratip}, therefore, we introduced derived types at various levels of
complexity, including for example a data structure \texttt{cs\_{}function}
for the representation of a single CSF (in $jj-$coupling), or the
derived type \texttt{cs\_{}basis} in order to 'store' the information about
a full configuration basis. To these data structures, memory is
allocated only on the basis of the actual requirements concerning the number of 
occupied shells and CSF in the given CSF basis. In \textsc{Lsj}, we now 
followed very similar lines and define the corresponding data structures also
in $LS-$coupling. Two simple examples are

\begin{scriptsize} 
\begin{verbatim} 
   type, public :: cs_function_LS 
      integer(kind=i1b) :: totalJ 
      character(len=1)  :: parity 
      integer(kind=i1b), dimension(:), pointer :: occupation 
      integer(kind=i1b), dimension(:), pointer :: seniority 
      integer(kind=i1b), dimension(:), pointer :: w 
      integer(kind=i1b), dimension(:), pointer :: shellL 
      integer(kind=i1b), dimension(:), pointer :: shellS 
      integer(kind=i1b), dimension(:), pointer :: shellLX 
      integer(kind=i1b), dimension(:), pointer :: shellSX 
   end type cs_function_LS 
\end{verbatim} 
\end{scriptsize} 

to represent a $LS-$coupled CSF in terms of the occupation and angular momentum
quantum numbers of all individual subshells as well as in terms of the
intermediate angular momenta $ L_{12}, S_{12},\, L_{123}, S_{123},\, ...$ which
are stored in the two arrays \texttt{shellLX} and \texttt{shellSX}. The data 
relating to all the $LS-$coupled CSF are then \textit{kept together} in the 
derived data type 

 \begin{scriptsize} 
\begin{verbatim} 
   type, public :: csf_basis_LS 
      integer :: nocsf         ! Number of CSFs in the basis. 
      integer :: nwshells      ! Number of (nonrelativistic) shells. 
      integer :: nwcore        ! Number of (closed) core shells. 
      integer :: number_of_electrons 
      type(nl), dimension(:), pointer             :: shell 
      type(cs_function_LS), dimension(:), pointer :: csf 
      type(parent_from_jj), dimension(:), pointer :: parent 
   end type csf_basis_LS 
\end{verbatim} 
\end{scriptsize} 

which, within the pointer array \texttt{parent}, also includes
details about the \textit{parentage} of the corresponding configuration states
functions in the original, $jj-$coupled CSF basis. Further derived structures
concern the representation of the atomic states (\texttt{asf\_function\_LS}
and \texttt{asf\_basis\_LS}) as well as a list of all $LS-$coupled subshell
states \texttt{subshell\_term\_LS} to identify the allowed terms for 
open--shell configurations. Obviously, the efficient use of such data types 
facilitates the access to and the handling of selected data. In the 
\textsc{Lsj} program, all new derived types are defined in the module 
\texttt{rabs\_lsj} (file \texttt{rabs\_lsj.f90}) which also contains the 
transformation coefficients (\ref{subshell-coefficients}) for the 
$ jj \,\leftrightarrow\, LS$ transformation of the subshell states. 
These matrix elements are kept in the derived structure

\begin{scriptsize}   
\begin{verbatim} 
type, private :: LS_jj_me 
      sequence 
      integer(kind=i1b)    :: w, Q, L, S, J, Nm, Qm, Jm, Qp, Jp  
      integer(kind=i1b)    :: factor 
      integer*8            :: nom, denom 
end type LS_jj_me 
\end{verbatim} 
\end{scriptsize} 

within a \textit{integer representation}, from which the value of the 
matrix element follows by computing simply:
${\tt factor} \sqrt{\frac{\tt nom}{\tt denom}}$.

{\renewcommand\baselinestretch{0.85} 
\begin{figure} 
\begin{small} 
\begin{verbatim} 
                 | 
               P |     |-------------------------------| 
               R |-----| lsj_initialization_LS_jj()   | 
               O |     |-------------------------------| 
               G | 
               R |     |-------------------------------| 
               A |-----| lsj_control_transformation() | 
               M |     |-------------------------------| 
                 |         | 
                 |         | 
               E |         |---  data input 
               X |         | 
               E |         | 
               C |         |---  formation of all CSF in LSJ coupling
               U |         | 
               T |         | 
               I |         |---  jj-LS transformation and classification 
               O |         |     of all selected levels
               N |         | 
                 |         | 
                 |         |---  data output 
                 |         
\end{verbatim} 
\end{small} 
\caption{\small Schematic flow chart of the \textsc{Lsj} program.} 
\label{fig:execution} 
\end{figure} 
}

\subsection{Interactive control and output of the program} 
 
Like the other components of \textsc{Ratip}, the \textsc{Lsj} program 
is controlled interactively. Figure \ref{fig:execution} displays the basic 
steps in which the program is executed. Following the initialization of
several data arrays at the beginning of the execution, the control is then 
taken by the procedure \texttt{lsj\_control\_transformation()}, which reads 
and interprets all input data, carries out the transformation for all selected
levels, and finally prints the results to screen. This procedure also prints
a \textit{full expansion} of all $jj-$coupled CSF in the corresponding 
$LS-$basis in the file \texttt{input-csl.LS}.

\medskip

Figure~\ref{fig:interactive} shows the interactive dialog of \textsc{Lsj}
which requests the user to provide the \textsc{Grasp92} configuration symmetry
list \texttt{(.csl)} file and the wave function representation (of the considered
level) in a corresponding mixing coefficient \texttt{(.mix)} file.
However, independent of the number of ASF which are represented in this 
\texttt{.mix} file, the user then needs to specify the level numbers
for which the $jj \,\rightarrow\, LS$ transformation is to be carried out, as
well as the number of leading $LS-$coupled CSF which are included in the 
printout. As discussed before, only the leading CSF is used for the $LSJ$ 
spectroscopic designation of the level but the 'weights' of the next most important
configuration states may still provide insight into the configuration mixing 
for the given atom or ion.
 
\medskip

As the transformation of atomic levels appears mainly as a \textit{geometrical
task}, the definition of the configuration basis along with the corresponding
mixing coefficients provide together all necessary information for a $LSJ$ 
classification of the atomic levels. For the transformation of the corresponding
wave functions, in contrast, some care must be taken also for the 
radial part of the wave function representation, for instance, by taking the
\textit{average} for each pair of radial orbitals with angular momentum 
quantum numbers $\kappa > 0$ and $-\kappa -1$, respectively, for all occupied
shells $(nl)$. Using the \textsc{Lsj} component, the transformation of the 
atomic levels is carried out \textit{promptly} for most open--shell structures;
the results are therefore printed immediately to screen. Moreover, to provide 
a test of having obtained a \textit{correct and complete} transformation,
the sum of all 'weights' in the $LS-$coupled basis, i.e.\ 
$\sum_{i} \, | c_i^{\,(LS)} (\tau) |^2$ (which, of course, should be 1), 
is also printed.

{\renewcommand\baselinestretch{0.91} 
\begin{figure} 
\begin{scriptsize} 
\begin{verbatim} 
 LSJ: Transformation of atomic states from a jj-coupled into a LS-coupled CSF basis (Fortran 95 version) 
  (C) Copyright by G Gaigalas, T Zalandauskas and S Fritzsche, Kassel and Vilnius (2003). 
   
 Transform one or several ASF from a GRASP92 calculation into a LS-coupled CSF basis.  
 The transformation starts from the given .csl and .mix files and is carried out  
 for the n leading CSF in the jj-coupled basis; the new representation in the LS basis  
 is printed similar as in a standard GRASP92 computation.         
   
 Enter the name of the GRASP92 configuration symmetry list file: 
> 
 Loading configuration symmetry list file ... 
  There are ... relativistic subshells; 
  there are ... relativistic CSFs; 
  ... load complete. 
 Enter the name of corresponding .mix mixing coefficient file: 
> 
 Loading mixing coefficients file ... 
  ... load complete; 
 Maximum number of considered ASF is:  
 Enter the level numbers of the ASF which are to be transformed, 
  e.g. 1 3 4  7 - 20  48  69 - 85 : 
> 
  Enter the number of leading CSF to be printed: 
> 
\end{verbatim} 

\vspace*{-0.5cm}
\end{scriptsize} 
\caption{\small Interactive dialog for \textsc{Lsj} at the beginning of the
                transformation.} 
\label{fig:interactive} 
\end{figure} 
}

\subsection{Distribution and installation of the program} 

Even though the main effort in developing the \textsc{Lsj} component 
certainly concerned the two (new) modules \texttt{rabs\_lsj} and
\texttt{rabs\_lsj\_data}, we re--distribute the \textsc{Ratip} package as a
whole, owing to the tight binding of the new components with the previously
developed code. Overall, therefore, the present version of \textsc{Ratip}
now contains 18 modules as well as the source code for the executables of the
seven components \textsc{Anco}, \textsc{Cesd99}, \textsc{Lsj}, \textsc{Rcfp},
\textsc{Relci}, \textsc{Reos99}, and \textsc{Utilities}. As previously, they are
all contained in the \texttt{ratip} root directory. This root also comprises
several makefiles for generating the executable as well as a number of test 
suites for a few components. From this root, the executables of the different
components are generated by running the command \texttt{make -f make-component},
that is \texttt{make -f make-lsj} in the present case.

\medskip

By utilizing a large set of different makefiles, it has been found convenient
to declare several global variables for the compilation and linkage of the 
program. These \textit{declarations} are now comprised within the script file
\texttt{make-environment} which avoids the user having to adopt each makefile
independently to the local architecture and computational environment. In fact,
the script \texttt{make-environment} only contains a very few lines but must be
\textit{source}d before, i.e.\ by typing \texttt{source make-environment},
to make this information available to the system. 
Using this set--up of the code,
\textsc{Ratip} has been found portable rather easily to different platforms 
such as \textsc{IBM RS}/6000, \textsc{Sun OS}, or to the \textsc{PC} world. 
Moreover, a \texttt{Read.me} file in the \texttt{ratip} root directory
contains further details to the installation.

\medskip
 
For most components of the \textsc{Ratip} package, we provide in addition 
a test suite in a subdirectory \texttt{test-component} of the root
where \texttt{component} refers to the names above. For example, the directory 
\texttt{test-lsj} comprises all four (input and output) files in order
to run and compare the two examples below. These tests are typically designed
to exhibit the capabilities of the program but may serve also as a quick test 
of the local installation.

{\renewcommand\baselinestretch{0.91} 
\begin{figure} 
\begin{scriptsize} 
\begin{verbatim} 
 .
 .
 Enter the name of the GRASP92 configuration symmetry list file: 
\end{verbatim} 
\vspace{-0.3cm}
$>$be-csl.inp 
\vspace{-0.3cm}
\begin{verbatim} 
 Loading configuration symmetry list file ... 
  There are  4  relativistic subshells; 
  there are  4  relativistic CSFs; 
  ... load complete. 
 Enter the name of corresponding .mix mixing coefficient file: 
\end{verbatim} 
\vspace{-0.3cm}
$>$be-relci.mix 
\vspace{-0.3cm}
\begin{verbatim} 
 Loading mixing coefficients file ... 
  ... load complete; 
 Maximum number of considered ASF is: 4 
 Enter the level numbers of the ASF which are to be transformed, 
  e.g. 1 3 4  7 - 20  48  69 - 85 : 
\end{verbatim} 
\vspace{-0.3cm}
$>$2 4 
\vspace{-0.3cm}
\begin{verbatim} 
  Enter the number of leading CSF to be printed: 
\end{verbatim} 
\vspace{-0.3cm}
$>$2 
\vspace{-0.3cm}
\begin{verbatim} 
 Weights of major contributors to ASF in jj-coupling: 
  Level  J Parity      CSF contributions 
    2    1   -      0.66673 of    2    0.33327 of    3 

 Definition of leading CSF: 
            2) 2s ( 1)   2p-( 1) 
                  1/2       1/2  
                               1    
  Total sum of   weights    is:   1.000000000000000 
  The new level is under the investigation. 
 Weights of major contributors to ASF in jj-coupling: 
  Level  J Parity      CSF contributions 
    4    1   -      0.66673 of    3    0.33327 of    2 
 
 Definition of leading CSF: 
            3) 2s ( 1)   2p ( 1) 
                  1/2       3/2  
                               1    
  Total sum of   weights    is:   1.000000000000000 
 
 Weights of major contributors to ASF in LS-coupling: 
  Level  J Parity      CSF contributions 
    2    1   -      1.00000 of    3 
                      Total sum of  weights  is: 0.9999999999999996 
    4    1   -      1.00000 of    2 
                      Total sum of  weights  is: 0.9999999999999991 
 
 Definition of leading CSF: 
         2)     2s( 1)  2p( 1) 
                   2S1     2P1     1P      1    
         3)     2s( 1)  2p( 1) 
                   2S1     2P1     3P      1    
 LSJ complete ... . 
\end{verbatim} 
\end{scriptsize} 
\caption{\small Dialog and output for determining the $LSJ$ notation of 
                low--lying levels of the $1s^2 2s2p$ configurations in 
                beryllium.} 
\label{fig:exam_one} 
\end{figure} 
}

\begin{table}\centering{ \textbf{
{\large Table I. Leading CSFs and corresponding weight coefficients 
        ($| c_r^{\,(jj)} (\tau) |^2$ and $| c_s^{\,(LS)} (\tau) |^2$)
        of the low--lying levels 
        of the $1s^2 2s2p$ configurations in beryllium.}}}\\
\vspace{5 mm}
\begin{small}
\begin{tabular}{ c | c || c | c}\hline \hline
\multicolumn{2}{l||}{} & \multicolumn{2}{l}{ } \\[-0.3cm]
\multicolumn{2}{c ||}{$\ketm{\Psi_\tau (J^P)}_{jj}$} & 
\multicolumn{2}{c}{$\ketm{\Psi_\tau (J^P)}_{LS}$} \\[0.2cm]
\hline \hline
\multicolumn{1}{l|}{} & \multicolumn{1}{l||}{}
& \multicolumn{1}{l|}{} & \\[-0.3cm]
   weights                  & CSFs &   weights                   & CSFs \\
$| c_r^{\,(jj)} (\tau) |^2$ &      & $| c_s^{\,(LS)} (\tau) |^2$ &      \\[0.2cm]
\hline 
&&&\\
     0.66673    & $\ketm{((2s^1)\frac{1}{2}, (2{p-}^1)\frac{1}{2}) ~ 1}$   & 1 & $\ketm{((2s^1){}^{2}S, (2{p}^1){}^{2}P) ~ {}^3P_{1}}$ \\
&&&\\
     0.33327    & $\ketm{((2s^1)\frac{1}{2}, (2p^1)\frac{3}{2}) ~ 1}$  &
  &   \\
\multicolumn{2}{l||}{} & \multicolumn{2}{l}{ } \\[-0.1cm]
\multicolumn{2}{l||}{$\sum_{i} \, | c_i^{\,(jj)} (\tau) |^2$ = 1} & 
\multicolumn{2}{l}{ $\sum_{i} \, | c_i^{\,(LS)} (\tau) |^2$ = 1}  \\[0.4cm]
%
\hline
&&&\\
     0.66673    & $\ketm{((2s^1)\frac{1}{2}, (2p^1)\frac{3}{2}) ~ 1}$   & 1 & $\ketm{((2s^1){}^{2}S, (2{p}^1){}^{2}P) ~ {}^1P_{1}}$ \\
&&&\\
     0.33327    & $\ketm{((2s^1)\frac{1}{2}, (2{p-}^1)\frac{1}{2}) ~ 1}$   &
  &   \\ 
\multicolumn{2}{l||}{} & \multicolumn{2}{l}{ } \\[-0.1cm]
\multicolumn{2}{l||}{ $\sum_{i} \, | c_i^{\,(jj)} (\tau) |^2$ = 1} & 
\multicolumn{2}{l}{ $\sum_{i} \, | c_i^{\,(LS)} (\tau) |^2$ = 1}  \\[0.4cm]
\hline\hline
\end{tabular}
\end{small}
\end{table}

\section{Examples} 

To explain the use and execution of the \textsc{lsj} component, we have 
selected two examples which concern (i) the well--known splitting of the 
beryllium terms into a singlet and triplet system and (ii) shows how the
program can be used for the classification of the lanthanides with an open
$f-$shell. In both cases, the configuration symmetry list and the corresponding
mixing coefficients have been generated by means of \textsc{Grasp92} 
\cite{Parpia/CFF/Grant:96} and the \textsc{Relci} component 
\cite{FritzscheFG:02} of \textsc{Ratip}, using simply a 
configuration interaction approach. The input and (expected) output of these 
two examples are provided also in the subdirectory \texttt{test-lsj}. 
In both cases, the transformation is performed very rapidly within less than 
a second of \textsc{cpu} time on a Pentium 166 MHz processor.

\medskip

For beryllium, the low--lying excited spectrum of the $1s^2 2s2p$ configuration 
consists of just four levels which belong to the two $^1P$ and $^3P$ terms
and have total angular momenta $J=1$ and $J = 0,1,2$, respectively. Since the
Hamiltonian matrix is diagonal in the total angular momentum $J$, only the two
levels with $J=1$ will exhibit a configuration mixing and, in fact, this
interaction is necessary to represent the two levels in terms of the 
$1s^{2}2s^{1}2{p\mbox{--}}^{1}$ and $1s^{2}2s^{1}2p^{1}$ relativistic
configurations. For the two levels with $J = 0$ and $J = 2$, 
the assignment of the $LSJ$ notation is trivial because there is each just 
one CSF with a weight coefficients equal to 1 in the expansion. 
But to specify also the 
$LSJ$ symmetry of the remaining two $J=1$ levels, we can run the \textsc{Lsj}
program as shown in  Figure~\ref{fig:exam_one}, where use is made of the 
configuration symmetry list \texttt{be-csl.inp} and the corresponding 
mixing coefficient file \texttt{be-relci.mix}.
As seen from the Figure~\ref{fig:exam_one} and Table I where 
the results of the transformation are summarized, 
the lower one of the $J=1$  levels (No.\ 2) has a 
$^3P$ symmetry, and the $^1P$ term is assigned to level No.\ 4. Moreover,
while a strong mixing of the configuration states obviously appears
in the $jj-$coupled basis, a (nearly) pure $LSJ$ symmetry is found for all 
four levels in a $LS-$coupled configuration basis.

\medskip

For simple shell structures, such as the low--lying excitation spectrum of
beryllium and for many other light elements, the $LSJ$ assignment can often 
be made just on the basis of the excitation energies (with respect to the 
ground state) and total angular momenta of the individual levels, if these data
are compared with experiment. However, such a procedure is not 
feasible if two or more open shells occur in some electron configuration, 
or if open $d-$ and $f-$shell elements are considered. In such cases, 
several tens or even hundreds of levels may belong to a single 
(nonrelativistic) configuration, making a $LSJ$ symmetry assignment on the 
grounds of the calculated energies very unreliable. As a typically example for
such a complex configuration, we consider the ground state of gadolinium 
(Z = 64), which has a [Xe] $4f^7 5d 6s^2$ configuration with 3106 possible
($jj-$coupled) CSF. To determine the $LS$ symmetries of the computed
ground state with its known total angular momentum $2^-$, we again
run the \textsc{Lsj} component using the two files \texttt{gd-csl.inp} and
\texttt{gd-relci.mix}, respectively. 
Figure~\ref{fig:exam_two} displays the
corresponding dialog and demonstrates that the ground state of 
gadolinium has a $^9D_2$ symmetry. 
The leading $LS$-- and $jj$-- coupled CSFs together with their weights 
(squares of mixing coefficients) are presented in Table II. 
The rather small mixing of other $LS$ terms 
shows, moreover, that the low--lying levels of most (neutral) medium and heavy 
elements are still well classified in terms of their $LSJ$ spectroscopy 
notation. Despite this classification, however, much larger (or even huge) 
wave function expansions are typically required in order to obtain accurate 
predictions of the level structure and properties of these systems.

{\renewcommand\baselinestretch{0.91} 
\begin{figure} 
\begin{scriptsize} 
\begin{verbatim} 
 .
 .
 Enter the name of the GRASP92 configuration symmetry list file: 
\end{verbatim} 
\vspace{-0.3cm}
$>$gd-csl.inp 
\vspace{-0.3cm}
\begin{verbatim} 
 Loading configuration symmetry list file ... 
  There are  22  relativistic subshells;
  there are  3106  relativistic CSFs;  
\end{verbatim} 
\vspace{-0.3cm}
$>$gd-relci.mix  
\vspace{-0.3cm}
\begin{verbatim} 
  ... load complete. 
 Enter the name of corresponding .mix mixing coefficient file: 
 
 Loading mixing coefficients file ... 
  ... load complete; 
 Maximum number of considered ASF is: 11 
 Enter the level numbers of the ASF which are to be transformed, 
  e.g. 1 3 4  7 - 20  48  69 - 85 : 
\end{verbatim} 
\vspace{-0.3cm}
$>$1 
\vspace{-0.3cm}
\begin{verbatim} 
  Enter the number of leading CSF to be printed:
\end{verbatim} 
\vspace{-0.3cm}
$>$30 
\vspace{-0.3cm}
%
%
\begin{verbatim} 
   Weights of major contributors to ASF in jj-coupling: 
  Level  J Parity      CSF contributions 
    1    2   -   0.23193 of 1403   0.21537 of  457   0.12535 of 2443   0.09234 of   57   0.05331 of  987 
 
 Definition of leading CSF: 
         1403) 4f-( 3)   4f ( 4)   5s ( 2)   5p-( 2)   5p ( 4)   5d-( 1)   6s ( 2) 
                  9/2       8                                       3/2            
                               7/2        7/2        7/2        7/2        2         2     
  Total sum of   weights    is:   0.9999999993812160 
   
 Weights of major contributors to ASF in LS-coupling: 
   
  Level  J Parity      CSF contributions 
  
    1    2   -      0.97616 of    2    0.01538 of    4    0.00483 of    6    
                    0.00297 of    1    0.00034 of    8    0.00005 of  100
                    0.00005 of   60    0.00004 of   98    0.00004 of   58
                    0.00003 of    3    0.00003 of    5    0.00002 of  102
                    0.00002 of   62    0.00001 of   15    0.00001 of   28
                      Total sum over  weight  is: 0.9999999993812226
  
 Definition of leading CSF:
         1)     4f( 7)   5s( 2)   5p( 6)   5d( 1)   6s( 2)
                   8S0      1S0      1S0      2D1      1S0      8S       8S       7D       7D       2   
         2)     4f( 7)   5s( 2)   5p( 6)   5d( 1)   6s( 2)
                   8S0      1S0      1S0      2D1      1S0      8S       8S       9D       9D       2   
         3)     4f( 7)   5s( 2)   5p( 6)   5d( 1)   6s( 2)
                   6P0      1S0      1S0      2D1      1S0      6P       6P       5P       5P       2   
         4)     4f( 7)   5s( 2)   5p( 6)   5d( 1)   6s( 2)
                   6P0      1S0      1S0      2D1      1S0      6P       6P       7P       7P       2   
         5)     4f( 7)   5s( 2)   5p( 6)   5d( 1)   6s( 2)
                   6P0      1S0      1S0      2D1      1S0      6P       6P       5D       5D       2   
         6)     4f( 7)   5s( 2)   5p( 6)   5d( 1)   6s( 2)
                   6P0      1S0      1S0      2D1      1S0      6P       6P       7D       7D       2   
         8)     4f( 7)   5s( 2)   5p( 6)   5d( 1)   6s( 2)
                   6P0      1S0      1S0      2D1      1S0      6P       6P       7F       7F       2   
        15)     4f( 7)   5s( 2)   5p( 6)   5d( 1)   6s( 2)
                   6D0      1S0      1S0      2D1      1S0      6D       6D       7F       7F       2   
        28)     4f( 7)   5s( 2)   5p( 6)   5d( 1)   6s( 2)
                   6G0      1S0      1S0      2D1      1S0      6G       6G       7D       7D       2   
        58)     4f( 7)   5s( 2)   5p( 6)   5d( 1)   6s( 2)
                   4D1      1S0      1S0      2D1      1S0      4D       4D       5S       5S       2   
        60)     4f( 7)   5s( 2)   5p( 6)   5d( 1)   6s( 2)
                   4D1      1S0      1S0      2D1      1S0      4D       4D       5P       5P       2   
        62)     4f( 7)   5s( 2)   5p( 6)   5d( 1)   6s( 2)
                   4D1      1S0      1S0      2D1      1S0      4D       4D       5D       5D       2   
        98)     4f( 7)   5s( 2)   5p( 6)   5d( 1)   6s( 2)
                   4D6      1S0      1S0      2D1      1S0      4D       4D       5S       5S       2   
       100)     4f( 7)   5s( 2)   5p( 6)   5d( 1)   6s( 2)
                   4D6      1S0      1S0      2D1      1S0      4D       4D       5P       5P       2   
       102)     4f( 7)   5s( 2)   5p( 6)   5d( 1)   6s( 2)
                   4D6      1S0      1S0      2D1      1S0      4D       4D       5D       5D       2   
 LSJ complete ... . 
\end{verbatim} 
\end{scriptsize} 
\caption{\small Dialog and output for determining the $LSJ$ notation for the
                ground state of gadolinium.} 
\label{fig:exam_two} 
\end{figure} 
}

\begin{table}\centering{ \textbf{
{\large Table II. Leading CSFs and corresponding weight coefficients
        ($| c_r^{\,(jj)} (\tau) |^2$ and $| c_s^{\,(LS)} (\tau) |^2$)
        of the ground state of gadolinium.}}}\\
\vspace{5 mm}
\begin{small}
\begin{tabular}{ c | c || c | c}\hline \hline
\multicolumn{2}{l||}{} & \multicolumn{2}{l}{ } \\[-0.3cm]
\multicolumn{2}{c ||}{$\ketm{\Psi_\tau (J^P)}_{jj}$} & 
\multicolumn{2}{c}{$\ketm{\Psi_\tau (J^P)}_{LS}$} \\[0.2cm]
\hline \hline
\multicolumn{1}{l|}{} & \multicolumn{1}{l||}{}
& \multicolumn{1}{l|}{} & \\[-0.3cm]
   weights                  & CSFs &   weights                   & CSFs \\
$| c_r^{\,(jj)} (\tau) |^2$ &      & $| c_s^{\,(LS)} (\tau) |^2$ &      \\[0.2cm]
\hline 
&&&\\
     0.23193    & $\ketm{(((4{f-}^3~\frac{9}{2},4{f}^4~8)\frac{7}{2},5{d-}^1~\frac{3}{2})~2}$   
                & 0.97616 & $\ketm{(4f^7~{}^8S,5d^1~{}^2D)~{}^9D_2}$ \\
&&&\\
                &      & 0.01538 &  $\ketm{(4f^7~{}^6P,5d^1~{}^2D)~{}^7P_2}$ \\ 
&&&\\
                &      & 0.00483 &  $\ketm{(4f^7~{}^6P,5d^1~{}^2D)~{}^7D_2}$ \\ 
&&&\\
                &      & 0.00297 &  $\ketm{(4f^7~{}^8S,5d^1~{}^2D)~{}^7D_2}$ \\ 
&&&\\
                &      & 0.00034 &  $\ketm{(4f^7~{}^6P,5d^1~{}^2D)~{}^7F_2}$ \\ 
&&&\\
                &      & 0.00005 &  $\ketm{(4f^7~{}^4D^6,5d^1~{}^2D)~{}^5P_2}$ \\ 
&&&\\
                &      & 0.00005 &  $\ketm{(4f^7~{}^4D^1,5d^1~{}^2D)~{}^5P_2}$ \\ 
&&&\\
                &      & 0.00004 &  $\ketm{(4f^7~{}^4D^6,5d^1~{}^2D)~{}^5S_2}$ \\ 
&&&\\
                &      & 0.00004 &  $\ketm{(4f^7~{}^4D^1,5d^1~{}^2D)~{}^5S_2}$ \\ 
&&&\\
                &      & 0.00003 &  $\ketm{(4f^7~{}^6P,5d^1~{}^2D)~{}^5P_2}$ \\ 
&&&\\
                &      & 0.00003 &  $\ketm{(4f^7~{}^6P,5d^1~{}^2D)~{}^5D_2}$ \\ 
&&&\\
                &      & 0.00002 &  $\ketm{(4f^7~{}^4D^6,5d^1~{}^2D)~{}^5D_2}$ \\ 
&&&\\
                &      & 0.00002 &  $\ketm{(4f^7~{}^4D^1,5d^1~{}^2D)~{}^5D_2}$ \\ 
&&&\\
                &      & 0.00001 &  $\ketm{(4f^7~{}^6D,5d^1~{}^2D)~{}^7F_2}$ \\ 
&&&\\
                &      & 0.00001 &  $\ketm{(4f^7~{}^6G,5d^1~{}^2D)~{}^7D_2}$ \\ 
\multicolumn{1}{l|}{} & \multicolumn{1}{l||}{} &\multicolumn{2}{l}{ } \\[-0.1cm]
& & 
\multicolumn{2}{l}{ $\sum_{i} \, | c_i^{\,(LS)} (\tau) |^2$ = 0.99998}  \\[0.4cm]
\hline\hline
\end{tabular}
\end{small}
\end{table}

\section{Conclusions and outlook} 

The new component \textsc{Lsj} of the \textsc{Ratip} package is presented for
deriving the $LSJ$ designation of atomic levels and wave functions from 
relativistic calculations. Such spectroscopic notation is needed not only
to analyze a great deal of spectroscopic data but also for the interpretation 
of many inner--shell phenomena such as the angular distributions of 
emitted Auger and photo electrons.
For $2p \longrightarrow 4s$ resonantly excited argon atoms, for example,
it was shown recently \cite{Ueda:01} that the electron emission can be 
understood properly only if the decay in the (two--step) cascade is 
treated coherently. However, while the atomic levels in the cascade can be 
specified experimentally rather easily owing to the existence of observed line spectra,
program tools such as \textsc{Lsj} are needed to understand the theoretical data
and to select those levels, which have to be involved in the computation.

\medskip

The design of \textsc{Lsj} within the framework \textsc{Ratip} enables 
the user to apply a $jj \rightarrow LS$ transformation for all atoms and
shell structures which can be calculated by means of \textsc{Grasp92} or any
other \textsc{Ratip} component. This includes, in particular, open $d-$ and 
$f-$shell elements which have attracted a lot of recent interest. For
atomic fermium with $ Z \,=\, 100$, for example, first measurements on the 
low--lying excitation spectrum have recently been carried out in Mainz and 
could be supported by means of extended theoretical computations 
\cite{Backe:01}. Therefore, we hope that the \textsc{Lsj} component might 
be of use also in improving the available date base on the lanthanide 
($Z \,=\, 57,\, ...,\, 71$) and actinide elements ($Z \,=\, 89,\, ...,\, 103$)
in the future.

\medskip

In the design and set--up of \textsc{Ratip}, our intention from the very 
beginning was to provide the user with an \textit{open environment} for 
his or her studies of atomic properties. In the last few years, therefore, 
various requests arose for the further development of the code concerning, for 
example, the radiative and dielectronic recombination of highly--charged ions, 
electron--impact processes or even the study of multi--photon excitation 
and ionization processes. Very likely, however, only a few of these requests 
can be fulfilled by us in the forthcoming years where our attention will first 
be directed to the radiative capture of electrons, i.e.\ the reversed process
to the photoionization. Therefore, in order to extend \textsc{Ratip} along 
additional lines, any support of coworkers or even the independent implementation 
of new components will be appreciated.

%
%
%
 
\vspace{0.5cm} 
{\renewcommand\baselinestretch{0.95} 
\begin{small} 
 
\end{small} } 
 

\begin{thebibliography}{000} 
 
\bibitem{Parpia/CFF/Grant:96} 
   F.~A.~Parpia, C.~F.~Fischer and I.~P.~Grant, 
   Comput.\ Phys.\ Commun.\ \textbf{94} (1996) 249. 
 
\bibitem{Fritzsche:01} 
   S.~Fritzsche, J.\ Elec.\ Spec.\ Rel.\ Phenom \textbf{114--116} (2001) 1155. 
 
\bibitem{Fritzsche/CFF/Dong:00} 
   S.~Fritzsche, C.~F.~Fischer and C.~Z.~Dong,  
   Comput.\ Phys.\ Commun.\ \textbf{124} (2000) 342. 
 
\bibitem{FritzscheFG:02} 
   S.~Fritzsche, C.~F.~Fischer and G.~Gaigalas, 
   Comput.\ Phys.\ Commun.\ \textbf{148} (2002) 103. 
 
\bibitem{SandarsB:65} 
   P.~G.~H.~Sandars and J.~Beck, 
   Proc.\ R.\ Soc.\ London\ \textbf{1289} (1965) 97. 
 
\bibitem{Armstrong:66} 
   L.~Armstrong, 
   J. Math. Phys. \textbf{7} (1966) 1891. 
 
\bibitem{Armstrong:68} 
   L.~Armstrong, 
   J. Math. Phys. \textbf{9} (1968) 1083. 
 
\bibitem{KarazijaJ:01} 
   R.~Karazija, V.~Jonauskas, 
   J. Math. Phys. \textbf{42} (2001) 5642. 

\bibitem{Gaigalas/RZ:HCI02} 
   G.~Gaigalas, Z.~Rudzikas and T.~Zalandauskas, 
    Book of Abstracts of XI HCI Conference, 1-6 September, 2002, 
    Caen, France, p.\ 48.

\bibitem{Gaigalas/ZR:02}
    G.~Gaigalas, T.~Zalandauskas and Z.~Rudzikas, 
    At.~Data  Nucl.~Data Tables, in press (2003).

\bibitem{Gaigalas/RF:97}
    G.~Gaigalas, Z.~Rudzikas and C.~Froese Fischer, 
    J.~Phys.~\textbf{B 30} (1997) 3747.

\bibitem{Gaigalas/RF:98} 
    G.~Gaigalas, Z.~Rudzikas and C.~Froese Fischer, 
    At.~Data  Nucl.~Data Tables, \textbf{70} (1998) 1. 

\bibitem{Gaigalas/SR:00}
   G.~Gaigalas, S.~Fritzsche and Z.~Rudzikas, 
   At.~Data  Nucl.~Data Tables, \textbf{76} (2000) 235. 
 
\bibitem{Gaigalas/Fri:02}  
   G.~Gaigalas and S.~Fritzsche,  
   Comp.\ Phys.\ Commun.\ \textbf{149} (2002) 39. 
 
\bibitem{Rudzika-book} 
   Z.~Rudzikas, {\sl Theoretical Atomic Spectroscopy} 
   (Cambridge Univ. Press, Cambridge, 1997). 
 
\bibitem{Gaigalas/Fritzsche/Fricke:01} 
   G.~Gaigalas, S.~Fritzsche and B. Fricke,  
   Comp.\ Phys.\ Commun.\ \textbf{135} (2001) 219. 
 
\bibitem{Fritzsche:97} 
   S.\ Fritzsche, Comp.\ Phys.\ Commun.\ \textbf{103} (1997) 51.   

\bibitem{Fritzsche:02} 
   S.~Fritzsche, Phys.\ Scr.\ \textbf{T100} (2002) 37.  
   
\bibitem{Ueda:01}
   K.~Ueda, Y.~Shimizu, H.~Chiba, M.~Kitajima, H.~Tanaka, 
   S.~Fritzsche and N.~M.~Kabachnik, J.~Phys.~\textbf{B 34} (2001) 107.

\bibitem{Backe:01}
   M. Sewtz, H. Backe, A. Dretzke, G. Kube, W. Lauth, P. Schwamb, 
   K. Eberhardt, C. Gr\"u{}ning, P. Th\"o{}rle, N. Trautmann, 
   P. Kunz, J. Lassen, G. Passler, C.~Z. Dong, S. Fritzsche and 
   R.~G. Haire,
   Phys.\ Rev.\ Lett., in press  (2003).
 
 

%
%
%
%
\end{thebibliography}
\end{document}